# Investigation of quasi-cleavage in a hydrogen charged maraging stainless steel

Jolan Bestautte[a], Szilvia Kalácska[a,b,*], Denis Béchet[c], Zacharie Obadia[d], Frederic Christien[a]

[a] Mines Saint-Etienne, Univ Lyon, CNRS, UMR 5307 LGF, Centre SMS, 158 cours Fauriel 42023 Saint-Étienne, France

[b] Empa, Swiss Federal Laboratories for Materials Science and Technology, Laboratory of Mechanics of Materials and Nanostructures, Feuerwerkerstrasse 39, Thun CH-3602, Switzerland

[c] Aubert & Duval, 63770 Les Ancizes France

[d] Airbus Commercial Aircraft, Toulouse, France

*Corresponding Author, E-mail: szilvia.kalacska@cnrs.fr

**Abstract**

Slow strain rates tests (SSRT) were conducted on hydrogen-containing specimens of PH13-8Mo maraging stainless steel. Hydrogen-assisted subcritical quasi-cleavage cracking was shown to take place during SSRT, thus accelerating material failure. Fractographic analysis showed that quasi-cleavage is composed of flat brittle areas and rougher areas. Using cross-sectional electron backscatter diffraction (EBSD) analysis of a secondary subcritically grown crack, we observed brittle cracks propagated across martensite blocks ahead the main crack tip. These cracks were stopped at high-angle boundaries. The crack direction was consistent with propagation along {100} type planes. High-resolution EBSD showed significant crystal lattice rotation, hence consequential plastic deformation, concentrated *between* the main crack tip and the cracks located ahead it. It is concluded that quasi-cleavage in the material investigated here consists of {100} cleavage cracks connected by ductile ridges. A discontinuous mechanism, involving re-initiation of new cleavage cracks ahead the main crack tip is suggested.

## Graphical Abstract

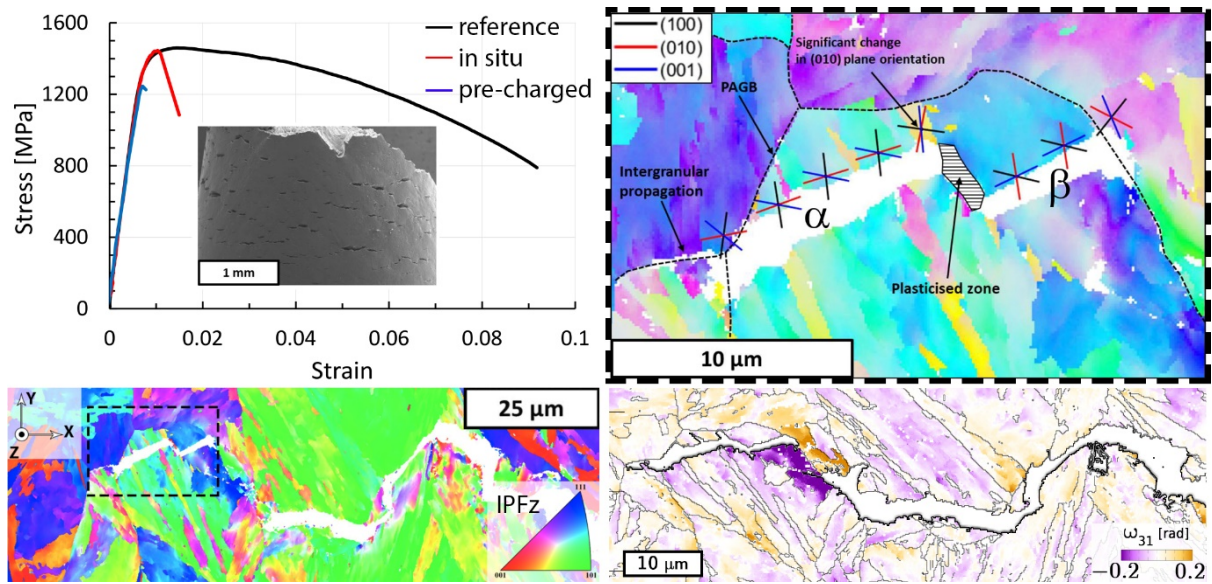

## Main text

Lath martensite in steels is a complex microstructure where former austenitic grains are divided in packets of blocks with different orientations [1]. Each block contains several slightly misoriented laths, all belonging to the same martensite variant (or to two slightly misoriented variants in case sub-blocks are present [1]). In most cases, lath or sub-block boundaries are low-angle, whereas block and packet boundaries are high-angle.

Martensitic steels are subject to hydrogen-assisted brittle fracture, that can be either intergranular or of "quasi-cleavage" type [2–7]. The term "quasi-cleavage" denotes a fracture mode where fine features, like steps and tear ridges, are present in or between cleavage-like facets [5,8]. Depending on studies, cleavage-like facets in H-charged martensite was shown to propagate along {110} [8–13] or {100} [5,6] lattice planes, the latter being the most expected cleavage plane type in BCC iron [14]. The propagation along {110} planes was interpreted as a separation along lath boundaries [8,10,13], because those interfaces are usually roughly aligned with the {110} habit plane of martensite, or as glide plane decohesion [11,12].

In a recent study, Cho et al. [5] showed that quasi-cleavage in a gaseous hydrogen-embrittled (HE) low alloy martensitic steel was a combination of flat cracks propagating mainly across the laths and along {100} planes, and rugged areas following no particular planes. They also pointed out the crucial role of boundaries in deflecting crack propagation.

The present study focuses on the PH13-8Mo maraging stainless steel, aged for 4 hours at 538 °C. Its yield strength reaches about 1400 MPa thanks to the nanoscale coherent precipitation of NiAl B2 phase [15] inside the martensite laths. A few percent of austenite is also present, mainly as nanometric films covering the different types of boundaries [16]. SSRT was performed under or after hydrogen cathodic charging. In contrast to previous works that mainly studied the main fracture surface [5,8,13], the focus in the current study was set on a secondary crack, i.e. not fully propagated across the specimen thickness, which allowed a detailed analysis of the crack tip region at a given instant of the propagation. This analysis was conducted using conventional electron backscatter diffraction (EBSD) on a cross-section obtained by broad beam ion milling. High (angular) resolution EBSD (HR-EBSD) [17, 18] evaluation of local lattice rotations and the dislocation density was also performed.

SSRT was carried out at multiple strain rates ($10^{-7} - 10^{-3}$ s$^{-1}$). Two H-charging methods were evaluated using i) samples that were charged during the tensile testing (referred to as "*in situ*") and ii) specimens that were pre-charged during 120 hours before deformation in air. More information about the charging and mechanical testing can be found in the Supplementary Materials.

Stress-strain ($\sigma - \varepsilon$) curves obtained at $10^{-6}$ s$^{-1}$ are shown in Fig. 1a. The yield strength and elongation at fracture for the reference specimen tested in air are ~1400 MPa and 9% respectively. It should be noted that the material shows very limited work hardening, as the ultimate tensile strength does not exceed 1430 MPa. The optical image of the reference sample after fracture (Fig. 1d) reveals significant necking, which is consistent with the progressive reduction in load visible on the tensile curve (black curve in Fig. 1a and b).

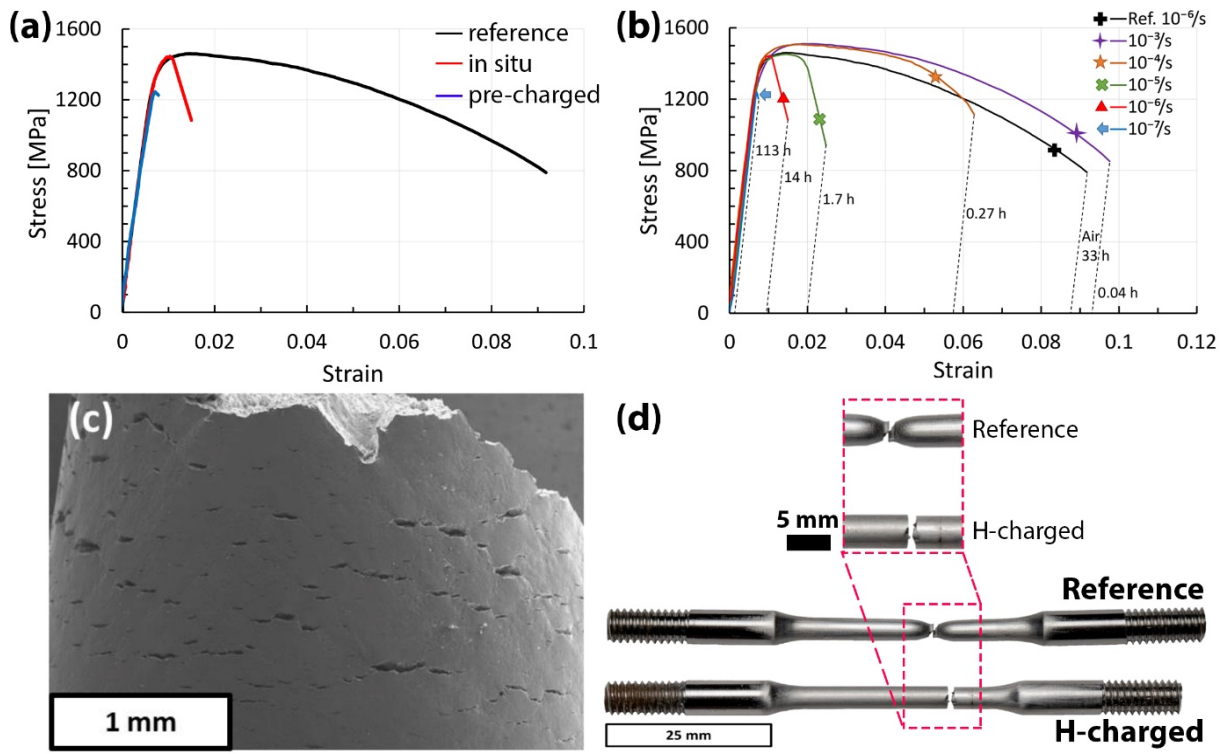

Fig.1. **Tensile test results.** (a) Stress-strain curves of PH13-8Mo ($10^{-6}$ s$^{-1}$ strain rate). The results obtained for tensile testing during cathodic charging (*in situ* condition, red curve) and for tensile testing in air after 120h of cathodic charging (pre-charged condition, blue curve) are compared to the reference test in air (black curve). (b) Influence of the strain rate on stress-strain curves during SSRT (*in situ* conditions). Note that higher flow stresses are observed at $10^{-3}$ s$^{-1}$ and $10^{-4}$ s$^{-1}$ experiments under H than at $10^{-6}$ s$^{-1}$ under air. This is most probably an effect of strain rate rather than an effect of hydrogen. (c) Secondary electron side view of the sample tested at $10^{-4}$ s$^{-1}$ with *in situ* H-charging, showing considerable secondary cracking. (d) Optical images taken of the reference and the *in situ* H-charged samples after SSRT at $10^{-6}$ s$^{-1}$.

Fig. 1a shows considerable reduction of the elongation at fracture at $10^{-6}$ s$^{-1}$ due to hydrogen in both *in situ* and pre-charged conditions. The optical image of the *in situ* H-charged sample after fracture (Fig. 1d) reveals no necking at all. The strain rate sensitivity of the *in situ* H-charged material was further investigated. In Fig. 1b, several *in situ* $\sigma - \varepsilon$ curves are simultaneously plotted, including the reference curve deformed at $10^{-6}$ s$^{-1}$ in air. The sensitivity to hydrogen embrittlement is shown to increase significantly with decreasing strain rate. Only at the strain rate of $10^{-3}$ s$^{-1}$, the H-charged sample could sustain similar elongation as the reference before failure. No necking was observed at strain rates ≤$10^{-5}$ s$^{-1}$, meaning that the material behaved purely in a brittle manner at the macroscale. The progressive reduction in the load visible on the tensile curves for the slowest strain rates is therefore not related to ductile necking. This supports the existence of subcritical cracking in the tensile specimens due to hydrogen. Fig. 1c is a side view of the specimen tested in *in situ* H-charging conditions at $10^{-4}$ s$^{-1}$, where a considerable number of secondary cracks are visible on the surface. This confirms the existence of H-assisted subcritical cracking, as already observed in similar materials in previous studies [2, 4].

Fig. 2 shows the fractographic observations conducted on the sample tested *in situ* at $10^{-6}$ s$^{-1}$. The overall fracture surface can be divided into three distinct regions. In the region closest to the edge of the sample, the crack presents a quasi-cleavage (QC) morphology (red area in Fig. 2a). The QC surface (Fig. 2d) consists of flat facets with river patterns and of other rougher areas. Fig. 2e is a higher magnification image of two flat facets (marked with white flags), clearly showing some river patterns, surrounded by rougher areas.

The second part of the crack corresponds to the green area in Fig. 2a and is located deeper in the sample. The fracture surface in this part of the crack appears to be different from the QC region, including more curved and embossed patterns (Fig. 2b). The existence of small dimples is presented in Fig. 2c, that are not apparent in the QC region. Moreover, cleavage rivers are hardly ever observed in this inner part of the crack. Therefore, due to the more ductile aspect, we refer to this crack morphology as ductile cracking (DC).

Looking at the morphologies of fracture surfaces obtained at a strain rate of ~$10^{-6}$ s$^{-1}$ (Fig. 2a-e), the following hypothesis can be made on the sequence that led to the failure of the sample. In a first stage (Fig. 2f), hydrogen-assisted cracks initiated and propagated subcritically under the effect of hydrogen towards the centre of the sample, following a QC propagation mode. In a second stage (Fig. 2g), one of the QC cracks reached a critical length for which the stress intensity factor at the crack tip exceeded the fracture toughness value of the material. Consequently, the crack propagation shifted from sub-critical to critical. The critical crack propagated at a very high rate, which leaves insufficient time for any hydrogen to diffuse and allow embrittlement to take place. This critical crack propagation is therefore most probably uncorrelated with the presence of hydrogen and corresponds to the DC morphology. In a third stage (Fig. 2h), the final failure of the sample was obtained by shearing of the remaining ligament, corresponding to the shear lips on the fracture surface.

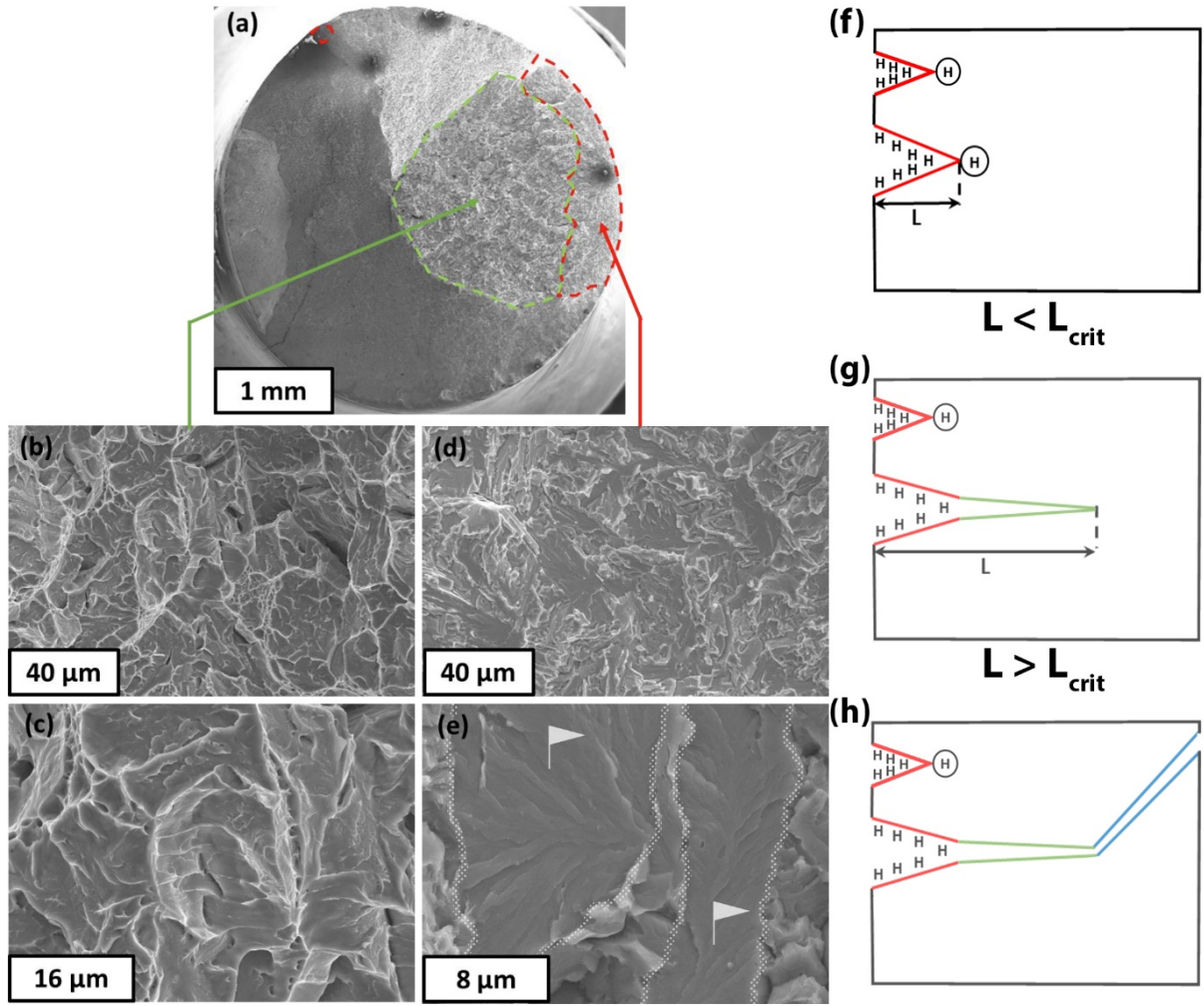

Fig. 2. **Fracture surface morphologies.** (a) Fracture surface of the sample tested *in situ* at $10^{-6}$ s$^{-1}$. The green region corresponds to ductile cracking (DC), associated with critical cracking. Higher magnification pictures of the DC region are presented in (b) and (c). The red region corresponds to quasi-cleavage (QC) cracking, associated with sub-critical cracking. Higher magnification pictures of the QC region are presented in (d) and (e). The rest of the fracture surface is composed of shear lips. (f-h) **Schematic representation of the sequence leading to the failure of a SSRT specimen in *in situ* H-charging condition.** (f) Slow propagation of a sub-critical crack assisted by hydrogen, which is adsorbed on crack surfaces and accumulates ahead of the crack tip. (g) Fast propagation of a critical crack after the sub-critical crack reached a critical length L$_{crit}$. The high crack growth rate most probably prevents any hydrogen/crack interaction. (h) Shearing leading to the final failure of the sample.

According to the scenario proposed above, all the secondary cracks have grown sub-critically but did not transition into critical cracks. It is then expected that secondary cracks consist of quasi-cleavage only. In order to deeper investigate quasi-cleavage, a cross-section of a secondary crack was prepared. For that purpose, the sample tested at $10^{-5}$ s$^{-1}$ with *in situ* H-charging was selected, as it showed significant secondary cracking, with no (or limited) generalized plasticity that could interfere with the EBSD measurements. More details about the sample preparation and the cross-section method can be found in the Suppl. Materials.

Fig. 3 shows the secondary electron observations and the EBSD analysis of the crack tip region. The tensile axis corresponds to the vertical direction. Note that only the crack tip region (located at 450 µm from the specimen edge) was studied. The crack has a segmented aspect and shows many bifurcations (Fig. 3a). It is noticeable that two small cracks (α and β) are present ahead the main crack tip. These two cracks are apparently disconnected from the main crack (however since we can only get information from a 2D section of a 3D fracture structure, the discontinuity cannot be fully supported by one measurement). The two small cracks are separated by a small, uncracked ligament. Each of these two cracks is relatively straight (no bifurcation is present). However, steps are clearly visible in Fig. 3b on the cracked surfaces (highlighted by white arrows). Defects are observed (black arrows) between the two small cracks (Fig. 3b) and in the uncracked ligament separating the two small cracks from the main crack tip (Fig. 3c). Furthermore, the later area appears darker in the EBSD band contrast map (Fig. 3d), which is a first indication of locally intense plastic deformation.

The band contrast and inverse pole figure (IPF) maps shown in Fig. 3d and e show that the cracks are transgranular, i.e. they do not seem to follow prior austenite grain boundaries (PAGBs), or any other type of boundaries. Fig. 3f is a higher magnification IPF map focusing on the two small cracks located ahead the main crack tip. The dotted lines show the PAGBs. Each crack goes through *one entire* martensite block. In other words, the cracks stop at PAGBs or block boundaries. A highly plasticized zone is present between the two cracks, where EBSD indexation rate was found to be lower. To assess if those cracks follow a specific crystallographic plane, the traces of {100} type planes are shown. Traces of {110}-type planes are also shown in Suppl. Fig. S4. The crack directions observed here are consistent with a propagation along {100} type planes. Suppl. Fig. S5 and S6 show complementary trace analysis of a secondary internal crack extracted from a pre-charged specimen tested at $10^{-5}$ s$^{-1}$.

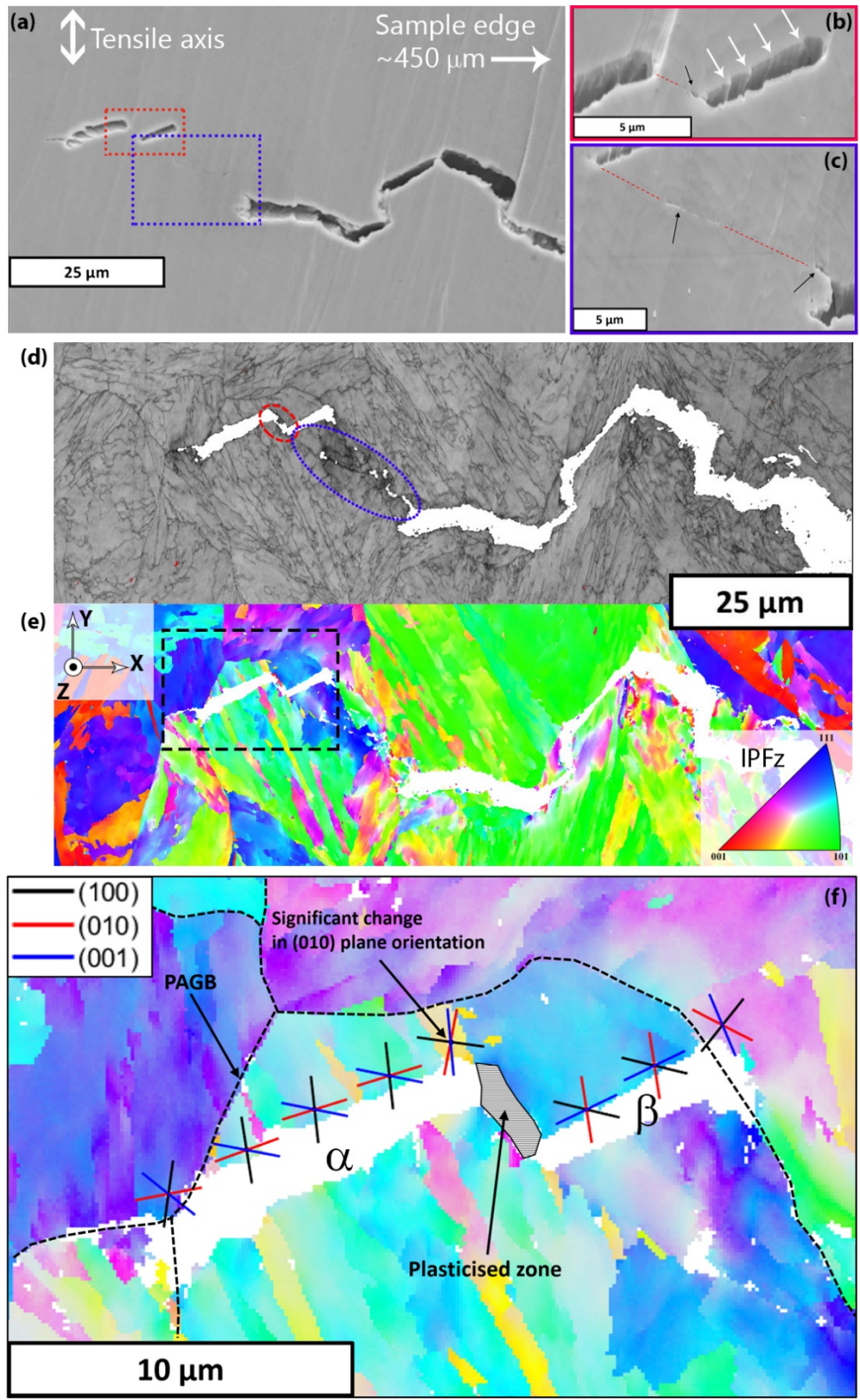

Fig.3. **EBSD results.** (a) Cross-sectional scanning electron micrograph of the secondary crack obtained *in situ* at 10$^{-5}$ s$^{-1}$. (b) Higher magnification of the two internal cracks. (c) Higher magnification of the ligament between the main crack tip and one of the internal cracks. (d) EBSD band contrast map. Low-indexation plasticised zones are highlighted with circles. (e) The corresponding IPF map. (f) Higher magnification IPF map corresponding to the area indicated in (e) with a black dashed rectangle. The traces of {100} type planes are indicated as well as the presence of plasticised zone.

Fig. 3d suggests intense plastic deformation concentrated between cracks. To study these plastic zones, HR-EBSD was applied to calculate the lattice rotation tensor elements ($\omega_{ij}$) and geometrically necessary dislocation (GND) density ($\rho_{GND}$) using the BLGVantage CrossCourt v4.0 software (Fig. 4). HR-EBSD is an image cross correlation technique utilized on the raw EBSD patterns [17], where a reference diffraction pattern is chosen at each grain to relate the state of each other pixel within that grain to this point. The results given by this technique are therefore relative to the reference patterns, that should ideally be selected at the minimal / zero stress state. In the absence of such references in our sample, HR-EBSD provide only a relative / qualitative support to the microstructure analysis to highlight the increased plastic zones around the cracks. More information about the HR-EBSD experiments are provided in the Suppl. Materials.

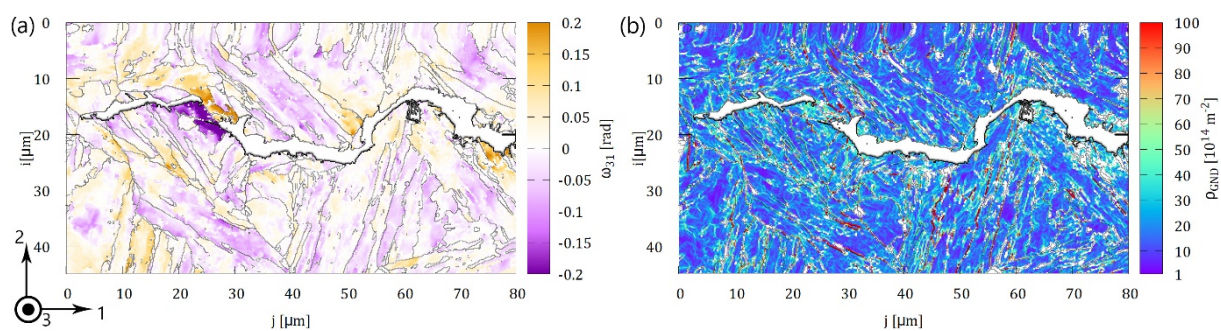

Fig.4. **HR-EBSD results.** (a) $\omega_{31}$ rotation tensor and (b) GND density map. Black lines highlight the high-angle grain boundaries (misorientation >10°).

Increased $\omega_{31}$ values (Fig. 4a) were found around the plasticised zone located between the main crack and the two internal cracks, indicating large lattice rotations in this area. This confirms the local intense plastic activity suggested by the band contrast map of Fig. 3d. The concentration of plastic deformation between the cracks suggests that this plastic deformation results from the mechanical interaction between cracks under tensile loading. This would imply that the cracks were formed first and the locally intense plastic deformation occurred later. Further components of the $\omega_{ij}$ tensor can be seen in Suppl. Fig. S7.

The estimated $\rho_{GND}$ map (Fig 4b) shows an overall high dislocation density (of the order of $10^{15}$ m²), comparable to other measurements in similar materials [19, 20]. The dislocations are mainly concentrated along lath boundaries, typical to martensitic steels [19]. On the other hand, no significant local increase of the dislocation density is measured in the plasticized zone located between the cracks. This can be related to the quasi-absence of work-hardening in this material, where plastic deformation is obtained from movements of the already existing dislocations, rather than by dislocation multiplication.

The investigations conducted here suggest a discontinuous mechanism of hydrogen-assisted sub-critical quasi-cleavage cracking, rather than a continuous propagation of the crack. High-angle grain boundaries are efficient obstacles to the crack propagation. These observations suggest that once the main crack tip is stopped at a high-angle boundary, new cleavage cracks of the size of one martensite block can form some micrometres (or even tens of micrometers) ahead of the main crack tip. This is possible most probably because of the high hydrostatic stress around the crack tip, that concentrates hydrogen there (note that the maximum hydrostatic stress is located at some distance ahead the crack tip, depending on the crack geometry). Once a new cleavage crack has formed, intense plastic deformation takes place in the ligament located between the main crack tip and the newly formed cleavage crack, until ductile failure of the ligament. In this mechanism, the local ductile failure is not expected to follow any particular crystallographic plane, as it is mainly governed by the stress field between the cracks. Once the newly formed cleavage cracks are connected to the main crack, the process can repeat itself further. The effect of hydrogen in this mechanism is of course expected in facilitating the formation of cleavage cracks, but it may also play a significant role in the failure of ductile connections.

Given this mechanism, quasi-cleavage in the material investigated here can be rationalised as a discontinuous process involving i) actual cleavage cracks propagated along {100} type planes across single martensite blocks, and ii) ductile connections between cleavage cracks. This leads to the aspect of quasi-cleavage shown in Figs 2d and 2e, where flat and rougher areas correspond to cleavage facets and ductile connections, respectively.

**Acknowledgements**

This research was mainly funded by Aubert & Duval and Airbus, with the support of ANRT (Association Nationale de la Recherche et de la Technologie).

Fruitful discussions in the CASHMERE R&T consortium (Airbus, Aubert & Duval, University of Manchester, Mines Saint-Etienne, La Rochelle University and French Corrosion Institute) are acknowledged, as well as the material supplied by Aubert & Duval.

The authors wish to express their sincere thanks to Claire Roume and Marilyne Mondon (Mines Saint-Etienne) for their technical support in this study.

**Declaration of competing interest**

The authors declare that they have no known competing financial interests or personal relationships that could have influenced the work reported in this paper.

# Investigation of quasi-cleavage in hydrogen charged aeronautical maraging stainless steel

Jolan Bestautte[a], Szilvia Kalácska[a,b,*], Denis Béchet[c], Zacharie Obadia[d], Frederic Christien[a]

[a] Mines Saint-Etienne, Univ Lyon, CNRS, UMR 5307 LGF, Centre SMS, 158 cours Fauriel 42023 Saint-Étienne, France

[b] Empa, Swiss Federal Laboratories for Materials Science and Technology, Laboratory of Mechanics of Materials and Nanostructures, Feuerwerkerstrasse 39, Thun CH-3602, Switzerland

[c] Aubert & Duval, 63770 Les Ancizes France

[d] Airbus Commercial Aircraft, Toulouse, France

*Corresponding Author, E-mail: szilvia.kalacska@cnrs.fr

**Charging and mechanical testing of the samples**

For *in situ* tests, the cathodic charging is performed by applying a potential of -1000 mV to a saturated calomel electrode (SCE). For *pre-charged tests*, the cathodic pre-charging is performed by applying the previously mentioned potential during 120h in the same solution and in the same cell. The cell is then removed, and the sample is dried with air before launching the SSRT test.

The polarization of the specimen during the electrochemical process was controlled by an external potentiostat (OrigaLys OrigaFlex). Three electrodes were present in the cell: i) a reference electrode as an SCE, ii) a counter electrode made of platinum mesh and iii) the sample acting as the working electrode. A deaerated liquid solution by nitrogen bubbling was continuously injected to the cell from a 3 liters external tank. The solution used for H-charging consisted of standard distilled water with 3.5 wt% NaCl at natural pH, kept at room temperature.

When performing hydrogen charging using cathodic polarisation, the deaeration of the solution is an important requirement. The reduction of dissolved oxygen interferes with the reduction of water, causing a decrease in hydrogen absorption inside the sample. In the case of the SSRT setup, a dissolved oxygen probe has been used to evaluate the efficiency of deaeration using nitrogen bubbling. The dissolved oxygen concentration as a function of deaeration time plot showed that the oxygen concentration reached a stable value of 8 ppb after 40 hours, therefore a deaeration time before launching the experiments was set as 48 hours.

Tensile deformation was carried out using a Schenck SCH-IN-1 instrument at various strain rates ($10^{-7} - 10^{-3}$ s$^{-1}$). The elongation of the specimens was recorded with an extensometer (reference sample) and two linear variable differential transformer (LVDT) sensors (charged samples).

The LVDT sensors give a less accurate measurement of the elongation than the extensometer, because the latter measures exactly the elongation of the sample gage length. However, the extensometer cannot be used inside the SSRT cell. For each material, a reference test in air was performed using both the extensometer and the LVDT sensors. The gap between the elongations measured by the extensometer and the LVDT sensors (only observed in the elastic region was used to correct the LVDT measurements. Eventually, the tests in *in situ* condition and in pre-charged-condition were performed using LVDT only, and the elastic region of strain-stress curves were corrected *post factum*.

**Sample preparation**

The materials were received as 100 mm sections of rods, which had either a rectangular section of 155 x 120 mm or a round section of Ø 90 mm in diameter. Then, electrical discharge machining was used to extract cylinders of Ø 7 mm in diameter and length of 100 mm in the longitudinal direction of the rods. Eventually, these cylinders were machined on a lathe to obtain the SSRT samples. The samples were polished using SiC paper down to P1200 grade, and then with diamond paste down to 1 µm grade. A lathe was used for polishing. Finally, the samples were rinsed with ethanol and distilled water, dried with air, and left in air during 24 h before testing.

**Residual stresses**

The residual stress inside the samples after machining can be significant. The residual stress close to the machined surface is usually high and can interact with hydrogen embrittlement. Especially, tensile residual stress close to the sample surface can promote the formation of cracks.

The polishing of samples has the dual purpose of reducing uncontrolled surface defects that can act as initiation sites for cracks and removing the material thickness affected by residual stresses. In order to assess the effect of polishing on residual stresses, measurements were conducted using X-ray diffraction (XRD) and electrochemical grinding in order to obtain a depth profile of residual stresses. These studies were performed on two SSRT specimens: one was in the as-machined state and the other one was polished after machining. The residual stresses were determined in the longitudinal and transverse directions. The spot size of these X-ray measurements is of the order of 1 mm in diameter. The results are presented in Suppl. Fig. 1. It can be seen that residual stresses are mainly found in the transverse direction, and are mainly located at the surface. It also appears that the polishing step had a very limited effect on the residual stress, probably because the layer of material removed during polishing is too thin (it was measured to be about 5 µm). However, the residual stresses appear to be mainly compression stresses in both machined and polished samples, which is supposed to mitigate cracks initiation. In addition, the stress level measured is far below the yield strength of the material in any case.

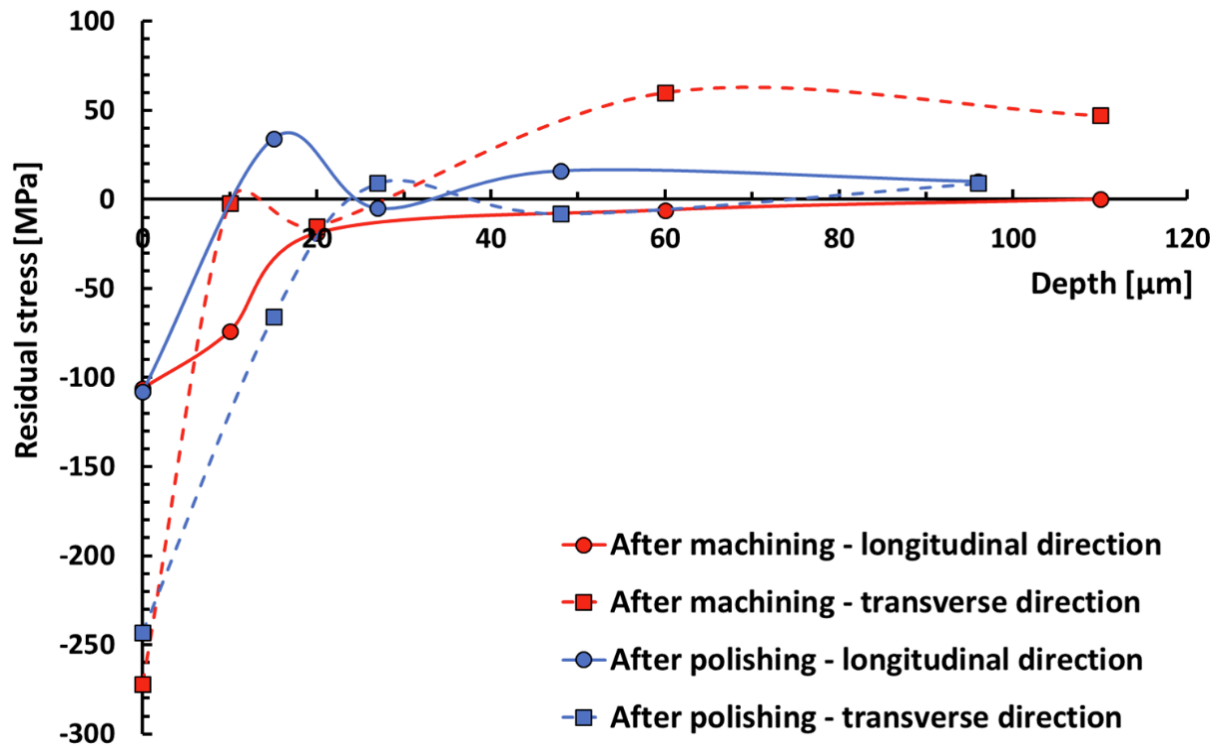

**Suppl. Fig. S1.** Residual stress measurements (by XRD) at the surface and subsurface of a SSRT sample after machining and after polishing.

**Post-deformation analysis**

A Zeiss Supra 55VP scanning electron microscope (SEM) equipped with an Oxford Instruments NordlysNano (EBSD) setup was used for fractographic observations and EBSD.

The secondary crack was extracted from the bulk sample by wire cutting, then Ar ion cross-polishing was applied on the surface (JEOL IB-19530CP).

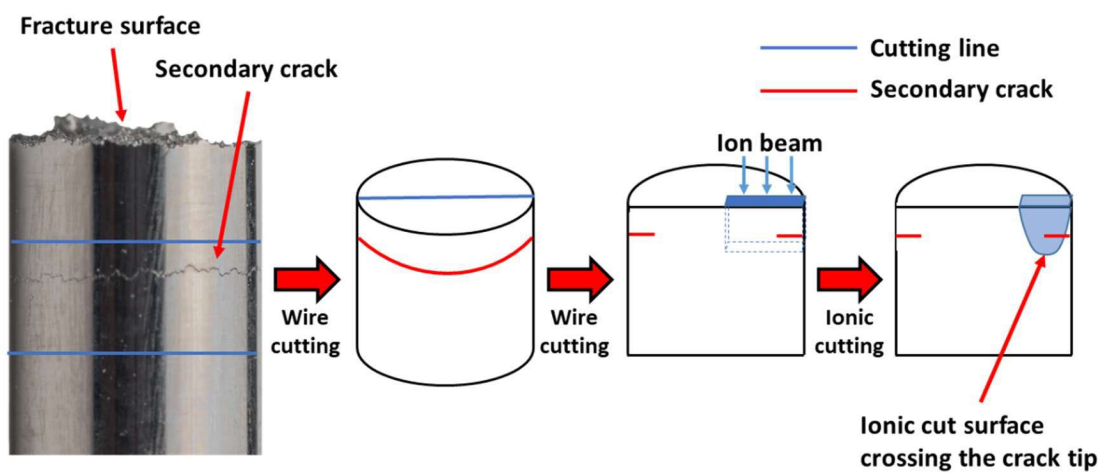

**Suppl. Fig. S2.** Schematic representation of the procedure used to extract and prepare the sample in order to study the crack path. The ionic cut surface, containing the crack tip can be directly analysed by EBSD.

Boundaries are marked with black lines in Suppl. Fig. S3, which correspond to PAGBs, packet and block boundaries. At first sight it does not seem that the crack is propagating along these boundaries. However, the lath boundaries cannot be clearly observed here because of their low misorientation. Nevertheless, it is known that inside a block, the lath boundaries are almost parallel with each other and with the block boundaries. The EBSD band contrast map in Fig. 3d seems to indicate that the two internal cracks are connected with each other, which is not the case, as previously shown in Fig. 1e-g. In the same manner, the connection between the internal cracks and the main crack tip seems more developed on the band contrast map than it really is according to the SEM observation. The reason is that the material is significantly deformed in these regions. It makes the EBSD indexation more difficult (due to blurred diffraction patterns), resulting in numerous not-indexed points that can be confused with effective cracks. The method used for the sample preparation prevented any material deformation. Therefore, it seems that it is the formation of the connections between cracks that involve significant plastic deformation.

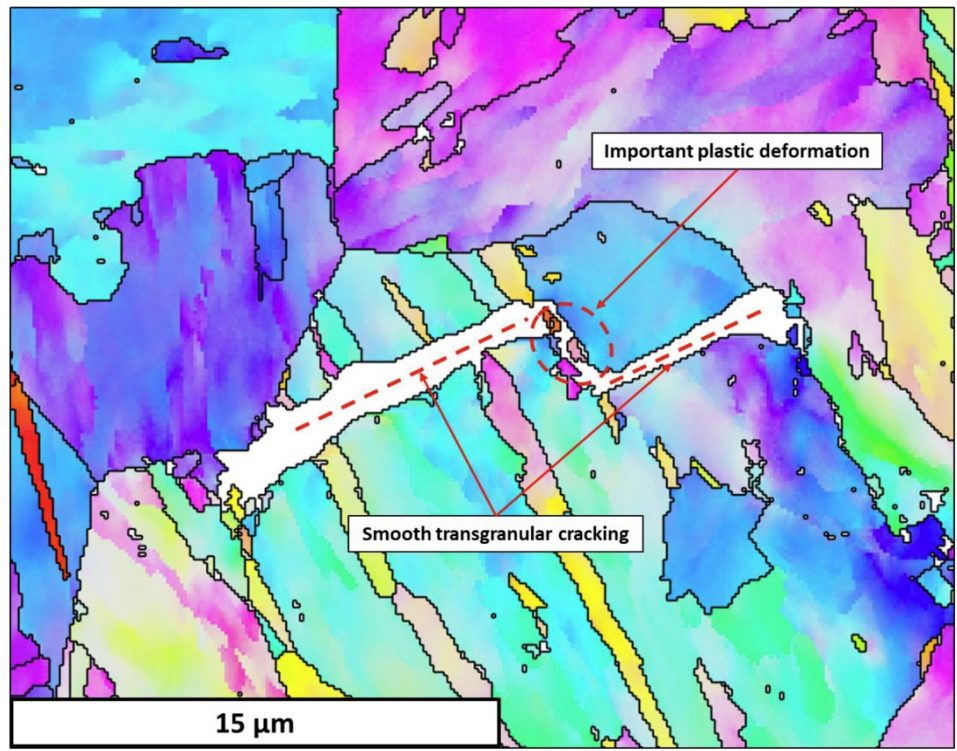

**Suppl. Fig. S3.** Magnification of EBSD IPF map on the internal cracks region. Detected boundaries (>3°) have been plotted with black lines. The two smooth sections seem to be transgranular. What seems to be a connection between the two smooth sections is in fact a heavily plasticised zone, resulting in low indexation.

The secondary phase austenite in maraging stainless steels mainly consists of thin layers of reverted austenite, located at the boundaries of the martensitic structure [S1, S2, S3]. However, it is evidenced here (Suppl. Fig. S3) that the QC cracks in this material do not follow grain boundaries. It is therefore not expected that the secondary austenitic phase interacts significantly with the hydrogen-assisted cracks.

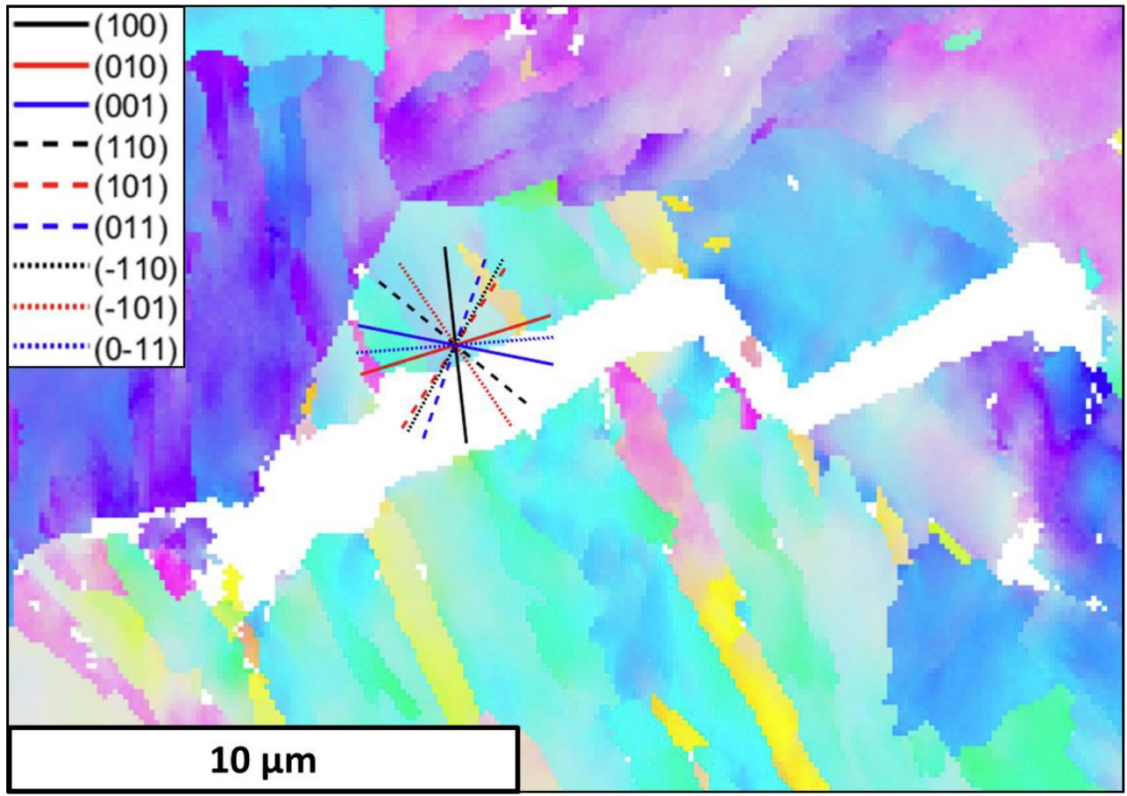

**Suppl. Fig. S4.** EBSD IPF map displaying local traces of the {100} and {110} planes close to the crack edge.

In Suppl. Fig. S4, the traces of {100} and {110} type planes are shown. The {100} planes are the cleavage planes in BCC iron, and the {110} planes are the martensitic laths habit planes. We remind the Reader that the martensitic matrix of the studied alloys presents a BCC structure (not tetragonal), mainly because of its very low carbon content. Thus, transgranular cleavage cracks should follow {100} planes and intergranular cracks that propagate along lath boundaries (or parallel block boundaries) should follow {110} planes. It can be seen that the only trace lying parallel to the crack edge corresponds to the (010) plane. As a consequence, it is likely that the crack locally follows a (010) cleavage plane.

**EBSD results on the deformed sample in pre-charged testing conditions**

In pre-charged condition, the cracks form internally, and rarely connect the side surface of the sample. As a consequence, it is challenging to cut through a crack, as it cannot be located before cutting. We only managed to cut through small secondary cracks, the largest one being investigated here, and we did not observe connections between cracks. Suppl. Fig. S5 presents a SEM observation and an EBSD band contrast map of a secondary crack after testing in pre-charged condition at a strain rate of $10^{-5}$ s$^{-1}$. According to the band contrast map, the internal crack presented here seems to follow a transgranular path.

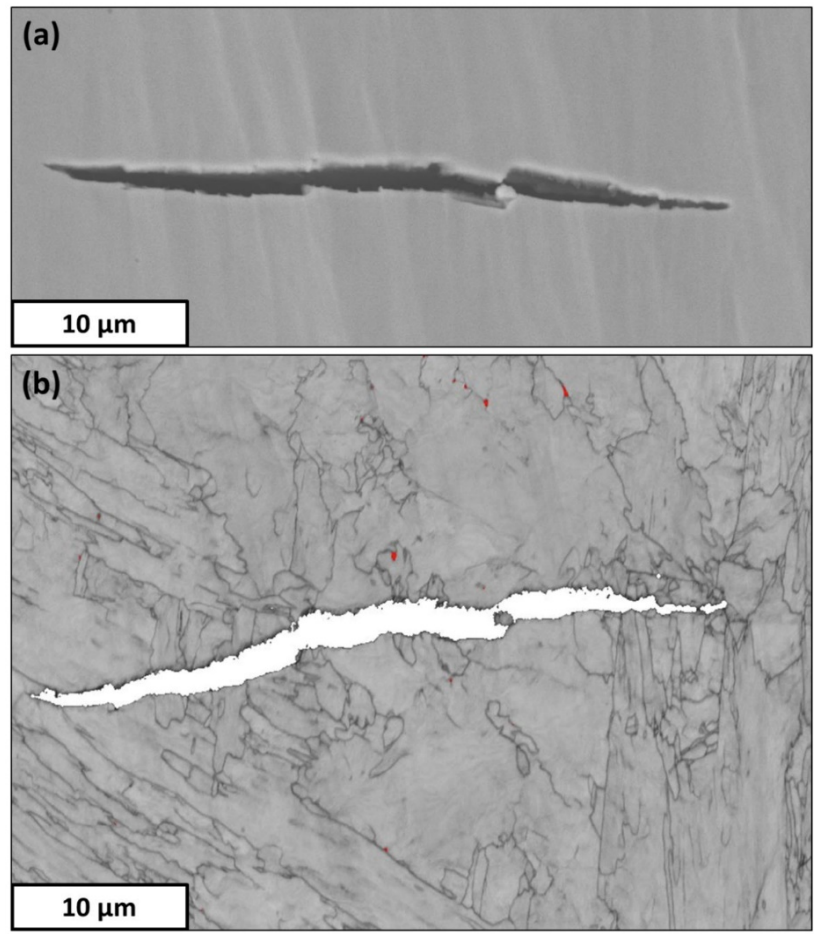

**Suppl. Fig. S5.** Internal crack in a sample tested at $10^{-5}$ s$^{-1}$ in H-pre-charged condition. (a) Cross-sectional SEM micrograph of the area selected for EBSD analysis and (b) EBSD band contrast map.

The traces of the {100} cleavage planes are presented in Suppl. Fig. S6a. It can be seen that the crack globally seems to follow {100} cleavage planes. This observation was made in the previous analysed cracks as well. However, traces of the {110} planes presented in Suppl. Fig. S6b show that some part of the crack better align with {110} planes.

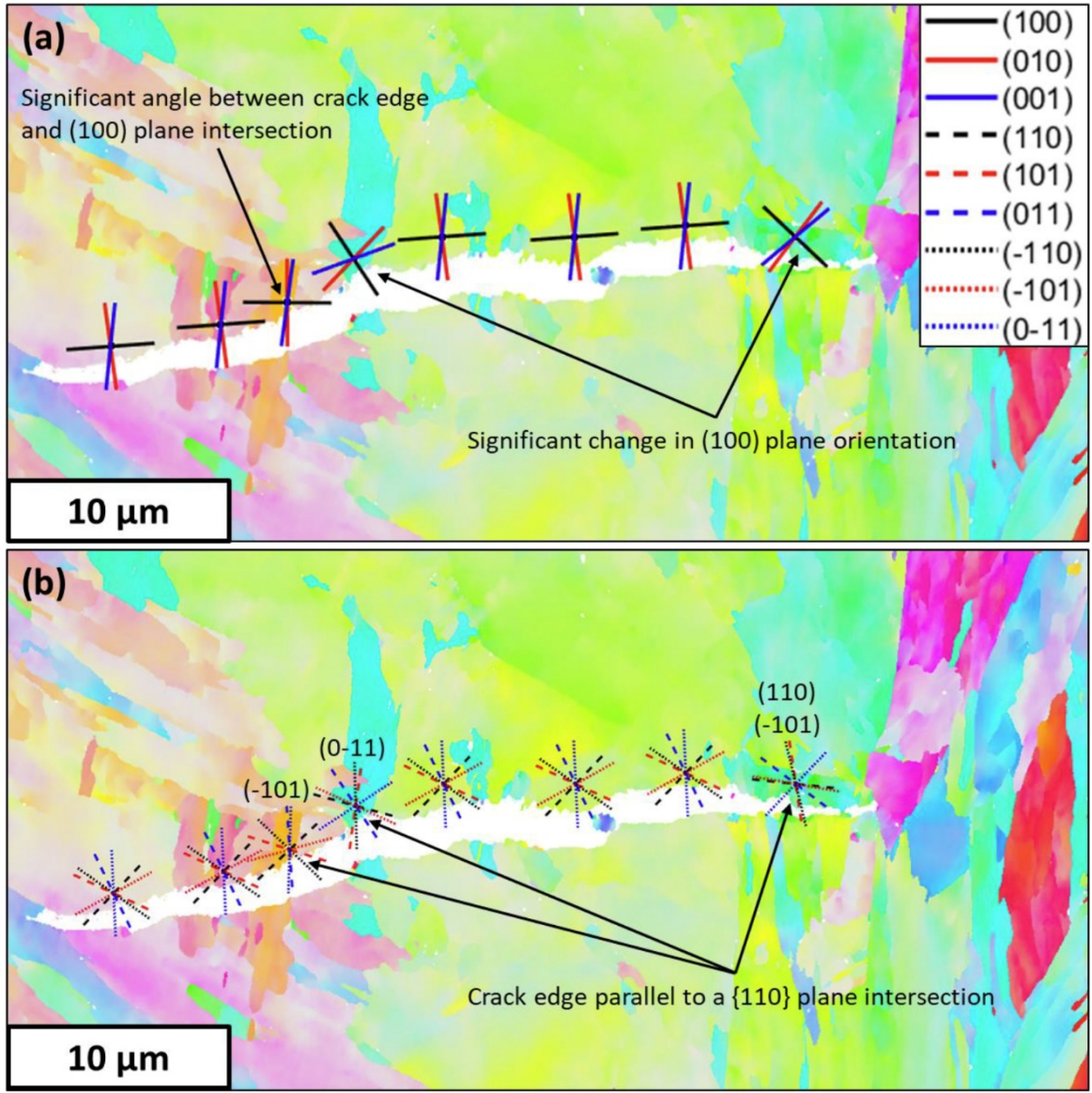

**Suppl. Fig. S6.** Internal crack in a sample tested at $10^{-5}$ s$^{-1}$ in H-pre-charged condition. EBSD IPF map showing traces of (a) {100} planes and (b) {110} planes. Overall, the crack seems to follow {100} cleavage planes, although better agreement is found with {110} planes in some part of the crack.

In summary, the QC cracking in pre-charged samples seems to consist in transgranular cleavage failure along {100} planes. It appears that the crack can also follow {110} planes, which are the habit planes of martensitic laths. Even though it suggests a failure along the lath boundaries, the EBSD map presented in Suppl. Fig. S6b indicates that the cracking occurred across the blocks. Therefore, it is unlikely that the crack followed lath boundaries, which are almost parallel to the block boundaries. However, the {110} planes are also slip planes in BCC iron, and transgranular hydrogen-assisted cracking along these planes has already been reported in the literature [12]. Thus, it appears that hydrogen-assisted cracking along {110} slip planes is possible in PH13-8Mo in the testing conditions of this study, despite not being the preferential cracking path.

Overall, the QC crack path in the microstructure is very similar between the samples tested in *in situ* and in pre-charged conditions. This observation is consistent with the fracture surface analyses, which revealed identical QC surface morphologies in these two testing conditions.

**HR-EBSD results**

For the HR-EBSD evaluation, the sample's surface was re-polished with Ar ions (3 kV, 15 min) to remove the oxidised layer that has formed after some storage time in air. The patterns were recorded with a 2×2 binning (672 × 512 px² resolution) with a step size of 150 nm and frame averaging of 6, using a 20 kV high current electron beam. Small grains containing less than 50 measurement points were excluded from the evaluation, along with the pixels with low quality diffraction patterns (i.e. data recorded at the crack). Fig. S7 shows the $\omega_{12}$ and $\omega_{23}$ rotation tensor components (the $\omega_{31}$ is shown in Fig.4).

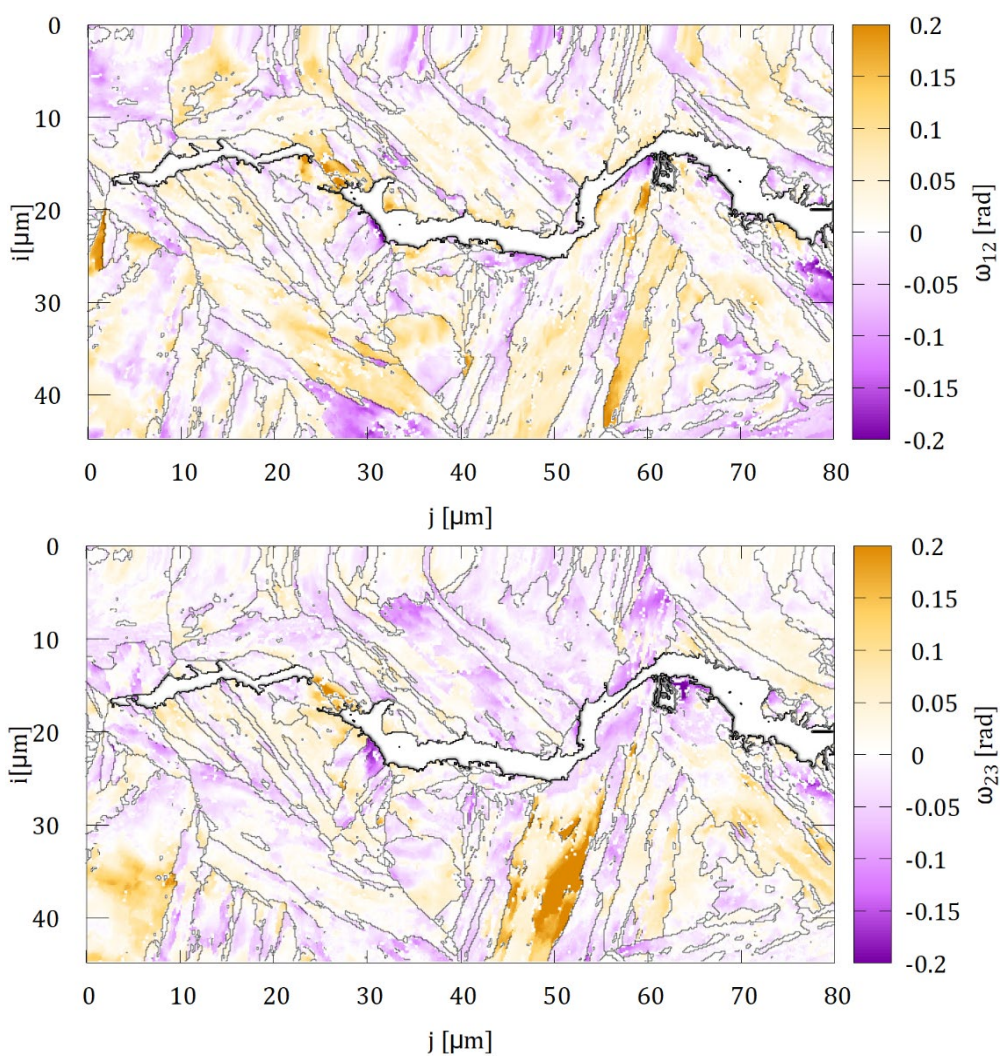

**Suppl. Fig. S7.** HR-EBSD rotation tensor components $\omega_{12}$ and $\omega_{23}$.

**Supplementary References**